\documentclass[preprint]{aastex}
\shorttitle{Bar of the LMC}
\shortauthors{A. Subramaniam \& S. Subramanian}
\begin{document}
\title{The Mysterious bar of the Large Magellanic Cloud: What is it?}
\author{
Annapurni Subramaniam\altaffilmark{1} \& Smitha Subramanian\altaffilmark{1}}
\affil{Indian Institute of Astrophysics, Koramangala II Block, Bangalore - 34}
\begin{abstract}
 The bar of the Large Magellanic Cloud (LMC) is one of the prominent, but
controversial feature regarding its location with respect to the disk
of the LMC. In order to study the relative location of the bar with respect to the disk,
we present the high resolution map of the structure across  
the LMC.  We used the reddening corrected mean magnitudes ($I_0$) of red clump (RC) stars
from the OGLE III catalogue to map the relative
variation in distance (vertical structure) or variation in RC population across the
LMC. The bar does not appear as an identifiable vertical feature in the map, as 
there is no difference in $I_0$ values between the bar and the disk regions.
We conclude that the LMC bar is very much
part of the disk, located in the plane of the disk (within 0.02 mag) and it is not a separate component.
We identify warps or variation in RC population with increase in radial distance.
\end{abstract}

\section{Introduction}
The off-centered stellar bar is one of the most striking features of the 
Large Magellanic Cloud (LMC). On the other hand, this is one of the least studied and
understood features of the LMC \cite{v06}.
Bars are a common phenomenon in late-type spirals and Magellanic irregulars
\cite{df73}.
The asymmetric bar is used to explain the recent star formation
history as well as the one-armed spiral feature in the LMC \cite{GTP98, DBCP96}.
\cite{A03} studied the relative distance within the LMC bar using red clump (RC) stars
and found that the bar is warped and also found extra planar structures in the bar. 
 \cite{Z04} suggested that the bar of the LMC is the result of
viewing a triaxial stellar bulge that is embedded in a highly obscuring thick disk.
The author also mentioned that, there are surprisingly few direct constraints 
on the three dimensional structure of this entity. Another reason to study the
vertical structure is to understand the results from the microlensing surveys.
\cite{ze00} proposed that this off-centered bar is an unvirialized structure slightly 
misaligned with, and offset from, the plane of the LMC disk. The small displacement 
and misalignment are consequences of recent tidal interactions with the Small
Magellanic Cloud (SMC) and the Galaxy.
Many considered this as a possible reason for microlensing events observed towards the LMC. 
\cite{N04} suggested the bar to be levitating above the disk by 0.5 kpc based on
distance estimations of Cepheids. Thus, location of the bar with respect to the disk is
still debated.
The near-IR star count maps presented by \cite{v01} found the bar to be a smooth
structure.
On the other hand, the bar is not visible in the HI distribution or in the HI
velocity maps \cite{sts03}.
Thus, when the bar is a prominent feature in the optical and near -IR, such a
feature is not
visible in the gas and recent star formation. Thus, the striking difference between
these two
distribution in the LMC has been an unsolved mystery. 

In this paper, we map the vertical structure (derived from the relative magnitude variation)
in the inner LMC using the recently published OGLE III catalogue \cite{u08}. 
Reddening corrected
mean magnitude of
RC distribution ($I_0$) is assumed to reflect the variation in distance along the line of
sight, such that,
regions with brighter peak magnitudes are assumed to be located closer.
This method was used by \cite{A03} and found a warped bar using the OGLE II data and
by \cite{os02}, who found a warp in the south-western disk.  Since the OGLE III scans
cover the disk region surrounding the bar as well as the bar, it is probably one of the best
data to study the relative location of the bar with respect to the inner disk. 
The high resolution map (64.5 x 64.5 sq.pc), presented in this study is expected to
bring out any difference in location, since the bar and the disk
are sampled. A difference in $I_0$ could also arise due to differences in RC population
between regions, due to variation in star formation history, metallicity etc.. Thus any
difference, or variation seen in the $I_0$ is interpreted as due to a vertical structure or
change in RC population. 

 The present analysis is expected to reveal the properties of the bar and the
disk as delineated by
RC stars, which are older than 1 Gyr. Thus the properties derived here pertain to the 
disk of intermediate age in the LMC. Since the bar of the LMC is believed to be
about 4 Gyr old \cite{Sh02},
RC stars are one of the best tracers to differentiate the bar from the disk.
The data is presented in the next section. Section 3 presents the results of the
vertical structure
in the LMC, followed by discussion in section 4.

\section{Data}
OGLE III survey \cite{u08} presented VI photometry of 40 square degrees of the LMC 
consisting of about 35 million stars.  We divided the observed region into 7169
regions 
(with reasonable no of
RC stars 100 - 2700) with a bin size of 4.44$\times$4.44 sq.arc min. The regions with
RC stars in the range 100 - 200 are located in the eastern and western ends of the
disk region covered by OGLE III. 
To obtain the number distribution of the RC stars,
they are binned in both colour and magnitude with a bin size of 0.01 and
0.025 mag respectively. These distributions are fitted with a Gaussian + Quadratic
polynomial. 
A non linear least square method is used for fitting and the parameters are
obtained. The parameters
obtained are the coefficients of each term in the function used to fit the profile,
error in
the estimation of each parameter and reduced chi square value. We estimated the
peaks in
I mag and (V$-$I) mag of the distribution, errors and goodness of fit.
Regions with peak errors greater than 0.1 mag and those 
with reduced chi square value greater than 2.0 are omitted from the
analysis. This rejection is similar to that adopted by \cite{A03}. 
The number of regions short listed to map the structure reduced to 5754.
 The analysis followed is similar to that presented in \cite{A03}.
In this analysis, we have not incorporated the incompleteness due to crowding, especially
in the central regions where the effect is expected to be prominent. In order to estimate
the effect due to crowding and the incompleteness, we compared the estimated parameters
with and without incompleteness correction \cite{SS09}. \cite{SS09} used OGLE II data for 
the analysis. Also, we compared the $I_0$ values
obtained from OGLE II data \cite{u00} incorporating incompleteness correction 
with the present estimates, in the next section.
We did not find any significant difference
between the parameters, suggesting that the incompleteness/crowding does not affect the
results presented here.

\section{Results: Structure of the inner LMC}
The peak values of the colour, (V$-$I) mag at each location is used to estimate the
reddening. The reddening is calculated using the relation
 E(V$-$I) = (V$-$I)$_(obs)$ $-$ 0.92 mag.
The intrinsic colour of the RC stars is assumed to be 0.92 mag \cite{os02}.
The interstellar extinction is estimated by $A_I$ = 1.4\,E(V$-$I) \cite{os02}.
After correcting the mean I mag for interstellar extinction, $I_0$ for each region 
is estimated. The difference in $I_0$ between regions is a measure of the 
relative distances such that $\sim$ 0.1 mag in $\Delta I$ corresponds to 2.3 Kpc in
distance.
The estimated high resolution map of the LMC is shown in figure 1. This shows the
variation in structure, which is basically the variation in relative distance in the
line of sight, and/or variation in RC population.
The center of the LMC is taken to be $05^h19^m38^s.0$ $-69^o27'5".2$ (2000.0)
\cite{df73}. 
 Only regions with total error in I$_0$ less than 0.03 mag are used for the
analysis and
a difference of 0.1 mag corresponds to a statistical significance of 3 $\sigma$.
We have also shown the approximate location of the bar, using two parallel lines.
This is the region covered  by OGLE II scans. Since same technique
was used by \cite{os02} for regions located mostly in the disk, we compared our
estimates
with their reddening corrected mean RC magnitudes. These are shown as red points in figure 1.
The points are shown such that bigger dots correspond to regions located closer (since they appear
brighter) and smaller dots correspond to regions located farther away (since they appear fainter).

The absence of any definite feature correlated to the location of the optical bar in the plot
is striking. 
This suggests that the bar is not located in front of the disk, at least in the
tracer adopted. Thus, the bar is likely to be very much part of the LMC disk. 
In order to compare the location 
of the disk studied by \cite{os02}, we have shown their locations in red, and the size of the dots follow the
same convention. OGLE III fields span a wider range in RA, but smaller range in Declination, whereas the regions
studied by \cite{os02} cover the Declination more, especially the northern disk regions.
It is should noted that this study has a continuous sampling of the central LMC, unlike \cite{os02}.
Their data match with the
present estimations very well in the overlapping regions. This is better verified in figure 3.

Another striking feature is the brightening of the RC stars towards the eastern end of the bar.
This region is outside the bar and located in the disk.
The line of nodes of the LMC is along the bar, therefore the 
magnitude variation due to the effect of inclination is not expected along the bar.
On the other hand,
the figure shows that regions located to the east of the bar along the same position
angle,
are brighter. Along the major axis, a mild warp in the western end is also suggestive. 
These features could also be due to a different RC population, perhaps due to a gradual change
in the star formation history or metallicity with radial distance. The warp  found by
\cite{os02} is in the regions which are located to the south-west of the OGLE III region.

 In order to see the effect of the reddening in the results arrived at, the
E(V$-$I) mean reddening
map is shown in figure 2. The minimum and maximum reddening values are 0.02 - 0.3
mag, where most
of the regions show reddening less than 0.15 mag. By comparing figures 1 and 2, it
can be seen that
the vertical structure, if any, as seen in figure 1 does not correlate with the pattern of
mean reddening.
Thus, the reddening correction does not have any bearing on the structure map
derived in figure 1.

 The edge on view of the LMC along the axes of minimum and maximum gradient are shown in figure 3. 
The line of nodes is taken as
PA = 121.$^o$5 \cite{v01}. The lower plot shows the edge on view, along the line of nodes and is not
expected to show any gradient in magnitude across the disk. The upper plot shows the magnitude variation
along the perpendicular axis, and this is the axis which shows maximum variation in magnitude. The direction
of the plotted axes are indicated.
The data from \cite{os02} are shown as red open circles. 
We have also
compared the $I_0$ estimates obtained by \cite{A03} for the OGLE II region, after incorporating
incompleteness correction, shown as green points.  These are shown only in the upper plot so that,
the lower plot is not crowded to hide any features suggesting a non-planar bar with respect to the disk.
The OGLE II data  agree very well with the OGLE III data. The data from \cite{os02} also match well. The
brighter points seen in the lower plot are from their north-eastern region, which are outside the OGLE III region. 
The bar is located between X= -2 to +2 degrees,
along with the disk in this edge on view.
In this central region, we see a single and relatively thin structure, suggestive
of the bar located within the disk and in the plane of the disk. 
Some regions located to the eastern and western ends
show signatures of warp, as suggested by the brighter points. 
 The upper panel is
suggestive of gradient in $I_0$, which is basically the inclination of the disk. We also find that the
north-eastern regions show relatively large range in I$_0$, when compared to the south-western regions.
The regions observed by \cite{os02} in the north-eastern side are located outside our field and are brighter, though
regions within OGLE III are comparable. Both the plots suggest that some regions located away from the center have 
brighter RC magnitudes, suggesting that there are extra-planar features/warps or variation in RC population, with radial
distance. 

In order to prove the coplanarity quantitatively,
we have averaged $I_0$ values in a bin of 1.$^o$5 in RA and 0.$^o$5 in Dec. Number of regions averaged vary between 25 -55. 
The bar and the nearby regions are
shown in figure 4. This plot also shows the rms error in the estimated mean within the bin. Thus the resolution shown in colour
code (0.02 mag) is of the order of 3 $\sigma$. Regions with error more than 0.01 mag are shown with crosses, where one
region has large error (0.03 mag). The averages shown for the bar are very similar to the result 
from OGLE II analysis \cite{A03}.
We also confirm their result of  eastern part of the bar located closer than the western end.
The average $I_0$ values in the bar region range between 18.12 to 18.18 mag (east to west), 
similar to the adjoining disk. The southern disk is brighter than the bar by 0.02 mag and
the eastern regions are brighter by 0.02 mag, suggesting that the bar is behind the eastern and the southern disk
by 0.02 mag. The western and the northern
disk have similar average $I_0$ as the bar, within 0.02 mag. There is one location in the northern region, where it is
fainter than the bar by 0.02 mag. 
The north-eastern regions have large rms errors due to less number of regions to average. 
These regions have $I_0$ values similar to the adjoining bar region. 
\cite{k09} found a probable brightening of
RC stars in the bar region by 0.04 mag. We do find the one localised region, near the center of the bar to show a similar brightening.
To summarise,  there is significant variation in the average value of $I_0$, along the bar as well as in the adjoining disk.
The bar and the disk are found to be coplanar within 0.02 mag.

\begin{figure}
\plotone{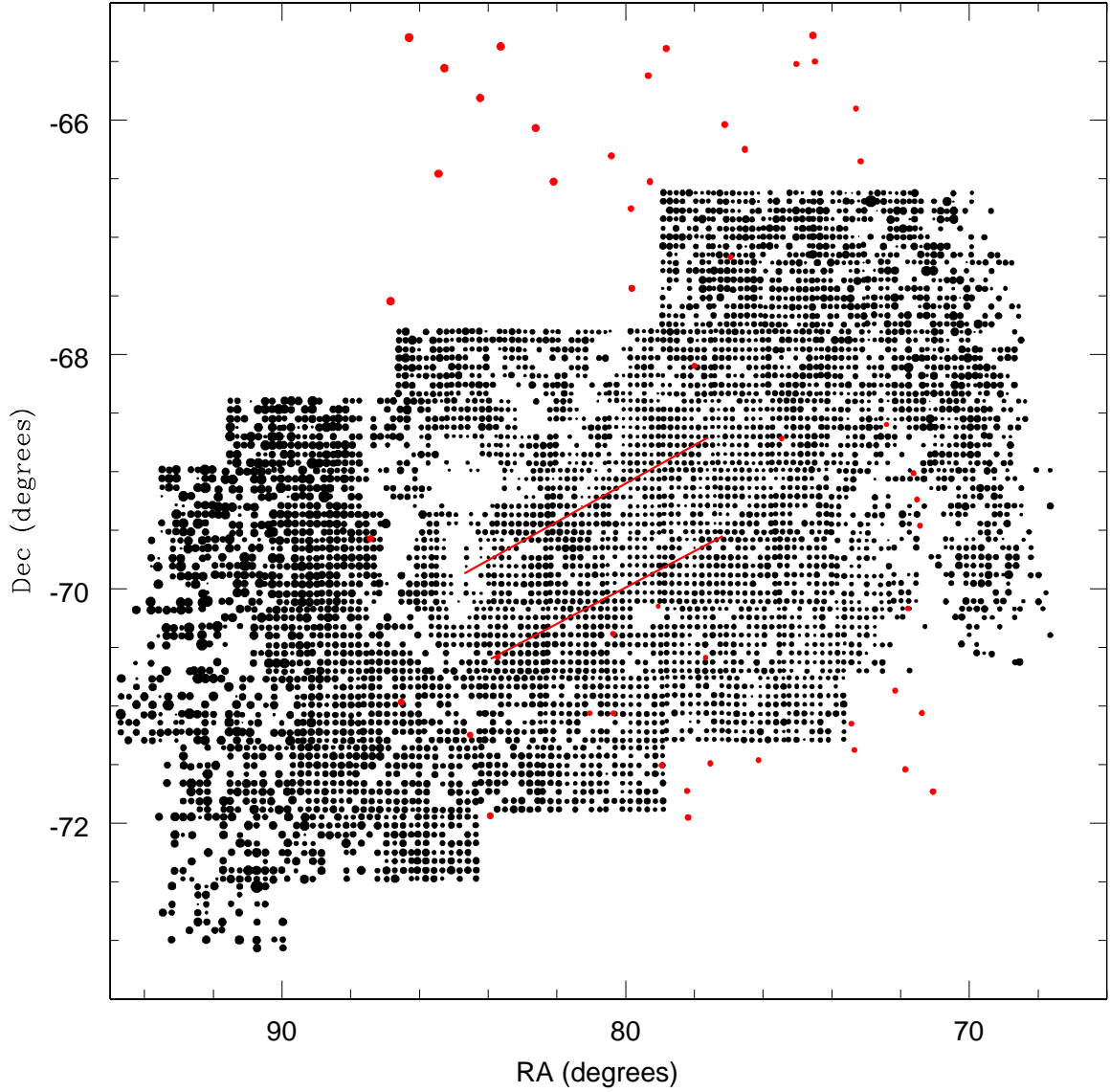}
\caption{Reddening corrected mean magnitude ($I_0$) of red clump stars estimated from OGLE III
data. The sizes
of dots are scaled such that the bigger points are brighter and smaller points are
fainter.
The range of $I_0$ shown here is from 17.9 - 18.3 mag (see figure 3). The red
parallel lines indicate
the approximate location of the bar. The red points correspond to data taken from
\cite{os02}.
\label{fig1}}
\end{figure}

\begin{figure}
\plotone{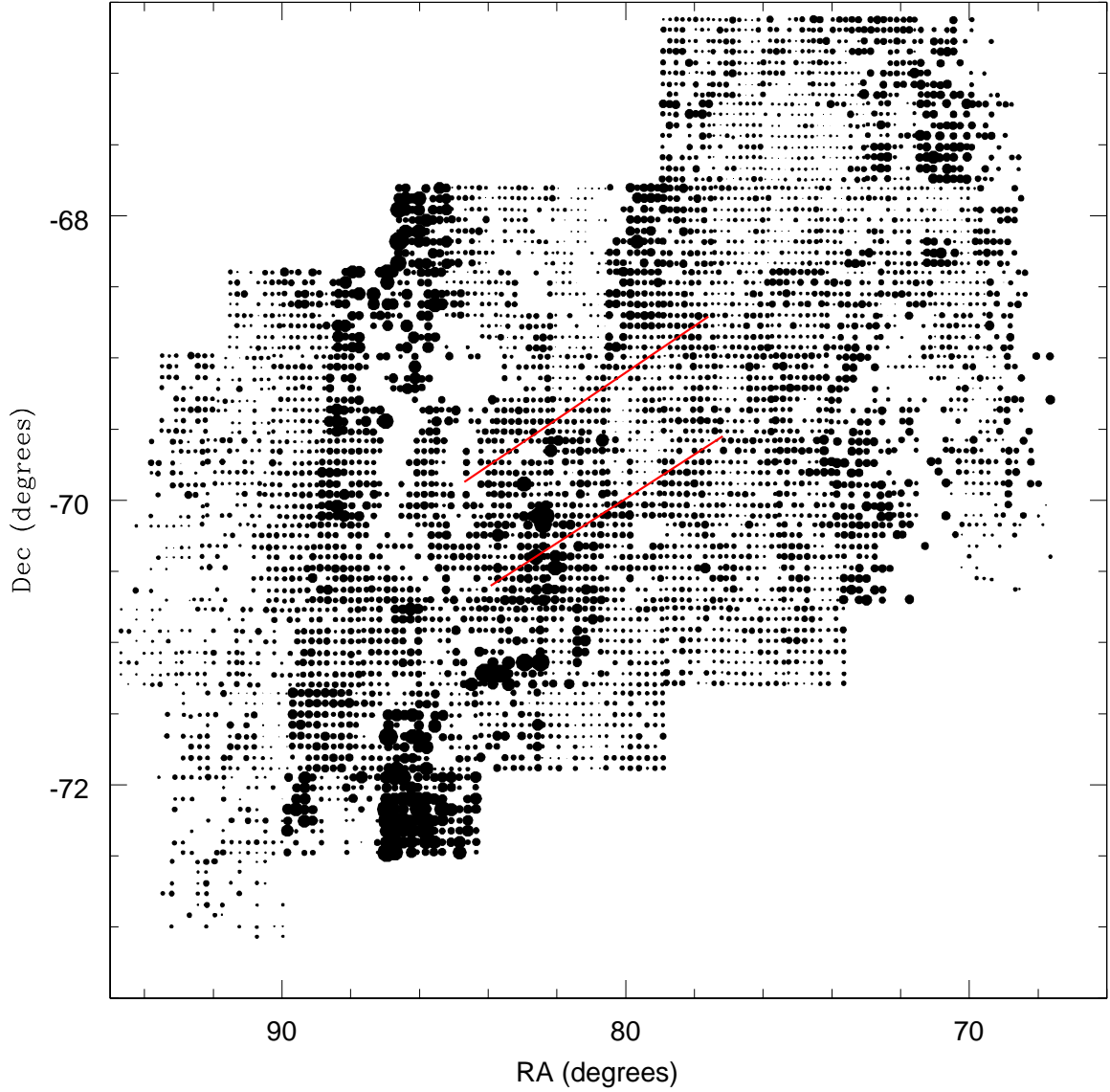}
\caption{ The mean reddening (E(V$-$I)) across the OGLE III field as estimated from
the red clump stars.
The sizes of the dots scale with the reddening value such that,
regions with large reddening (0.2 - 0.3 ) mag appear as bigger dots. Approximate
location of the bar
is indicated by the red parallel lines.
\label{fig2}}
\end{figure}

\begin{figure}
\plotone{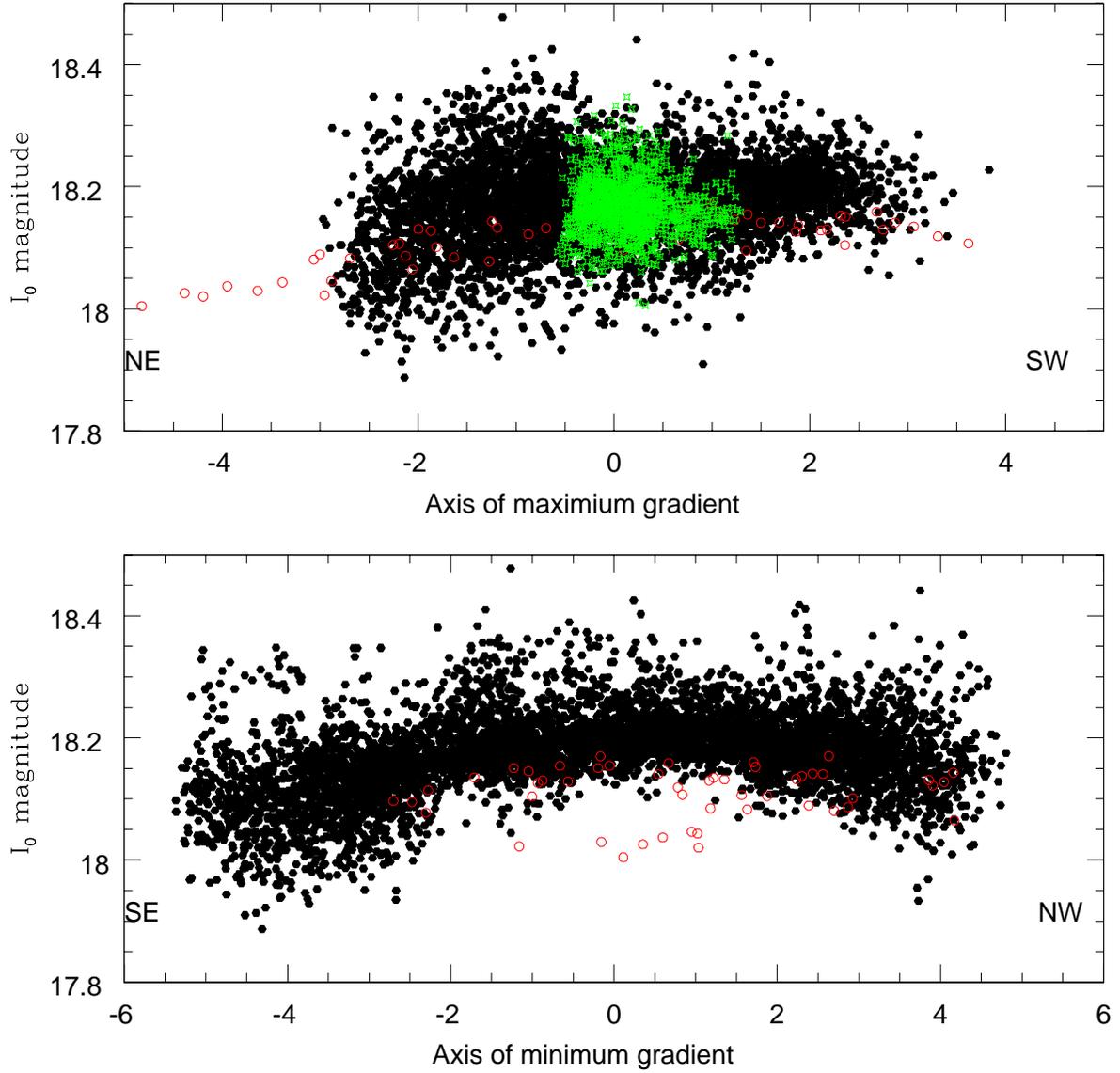}
\caption{The $I_0$ values are plotted along the axis of minimum gradient (line of nodes, PA = 121.$^o$5, lower panel)
and
along the axis of maximum gradient (perpendicular to the line of nodes, upper panel). 
The red open circles correspond to the data taken from
\cite{os02} and green points
correspond to data taken from \cite{A03}. 
\label{fig3}}
\end{figure}

\begin{figure}
\plotone{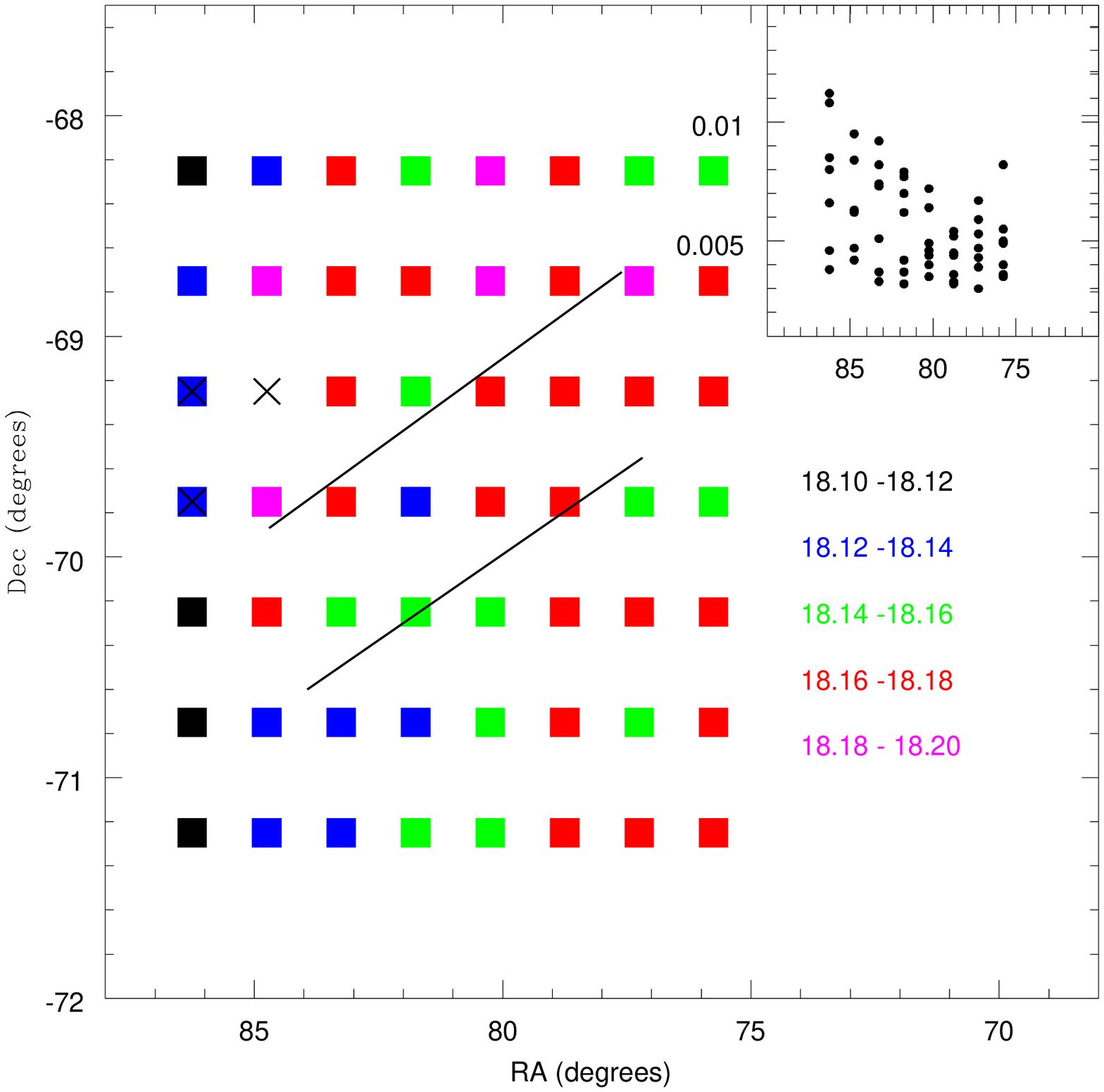}
\caption{The average values of $I_0$ near the bar region.
The rms errors weighted with the number of regions used are shown in the inset.
 The color code used and the bar region are shown.
The region with crosses over plotted are those with error more than 0.01 mag. The region with only cross shown is
where the error is very large (0.03 mag).
\label{fig4}}
\end{figure}

\section{Discussion}
The result derived here rules out the possibility that the bar may be located in
front of the disk.
The reddening in the LMC has been found to vary
with the tracer, 
\cite{A05} \&  \cite{Z99} and hence same tracer is used for reddening estimation. Also, the data
presented here
agree very well with the previous estimations by \cite{A03} and \cite{os02}. The
analysis of \cite{os02}
had points which are widely separated and \cite{A03} used only the bar region. Thus
a comparative study
of the bar and the disk with homogeneous data was not possible so far. The effect of
crowding on the
parameter estimation while studying the bar and the outer disk regions was
considered by \cite{SS09}.
They found marginal difference when one included incompleteness correction for the
bar region.
Thus the study presented here uses a method which was used earlier and uses a
homogeneous data which cover the bar and the inner disk.

The structure map clearly shows that the LMC bar is very much part of the LMC disk
and cannot be differentiated from the disk as a separate entity.
The estimated value of $I_0$ shows similar range in the bar as well as in the disk
region, making it difficult to differentiate the bar from the disk. 
The observed range of $I_0$ average values (18.12 - 18.18 mag) in the 
bar region, tallies very well with the surrounding disk within 0.02 mag.
\cite{ze00} suggested that the bar could be a misaligned and off-set along the line
of sight direction
by about 2 kpc. This study indicates that there is no difference between the mean
distance
to the bar and disk, though both the entities have large dispersion about the mean,
indicating a not so thin disk and the bar \cite{SS09}.

The range of estimated $I_0$ value decreases for regions located at the
eastern and western end of the observed region. The figures also suggest a warp in
the eastern and 
the western end of the observed region along the major axis. This is in addition to
the warp found by
\cite{os02} in the south-western part of the LMC disk. The gradual brightening of
the RC stars with 
increasing radial distance from the center could also be due to change in the RC
population.
A detailed analysis of the star formation history along with metallicity is required to
identify whether these are real warps or just different RC population.



\end{document}